\documentclass[twocolumn,showpacs,preprintnumbers,amsmath,amssymb,prb]{revtex4}
\usepackage{graphicx}
\usepackage{dcolumn}
\usepackage{bm}
\begin{document}
\title{Emergence of non-Fermi-liquid behavior due to Fermi surface
reconstruction
in the underdoped cuprate superconductors}
\author{Tanmoy Das, R. S. Markiewicz, and A. Bansil}
\address{Physics Department, Northeastern University, Boston MA 02115,
USA}
\date{\today}
\begin{abstract}
We present an intermediate coupling scenario together with a model
analytic solution where the non-Fermi-liquid behavior in the underdoped
cuprates emerges through the mechanism of Fermi surface (FS)
reconstruction. Even though the fluctuation spectrum remains nearly
isotropic, FS reconstruction driven by a density wave order breaks the
lattice symmetry and induces a strong momentum dependence in the
self-energy. As the doping is reduced to half-filling, we find that quasiparticle (QP) dispersion
becomes essentially unrenormalized, but in sharp contrast the QP spectral
weight renormalizes to nearly zero. This opposite doping evolution of the
renormalization factors for QP dispersion and
spectral weight conspires in such a way that the specific
heat remains Fermi liquid like at all dopings in accord with experiments.
\end{abstract}
\maketitle\narrowtext
\section{introduction}

Understanding how `non-Fermi-liquid' behavior arises near the half-filled
insulating state is one of the key questions for unraveling the physics of
not only the cuprates but that of correlated electron systems more
generally.  Here we show how some aspects of the electronic spectrum which
are difficult to understand in a conventional Fermi liquid theory can be
explained naturally in a model of an antiferromagnetic Fermi liquid.
Specifically, it has been reported that the renormalization of the
electronic dispersion decreases with underdoping as half-filling is
approached,\cite{sahrakorpi} i.e. the quasiparticles (QPs) seem to
`undress' in the underdoped regime. In sharp contrast, the spectral weight
of the QPs fades away on approaching the insulator and renormalizes to
zero at half-filling\cite{Yos06,yoshida}, indicating that these QPs are
very fragile or `gossamer'-like\cite{goss}.  Within Fermi liquid theory
the QP dispersion and spectral weight should be renormalized by the same
factor, unless the self-energy has a strong momentum dependence.
However, fluctuations in the cuprates seem to be relatively
isotropic\cite{water3,jarrell}, which would imply then that the
renormalization factor $Z_d$ for dispersion is roughly equal to the
renormalization factor $Z_{\omega}$ for the spectral weight. This is
clearly violated near the Mott insulating limit, where $Z_d\rightarrow 1$
while $Z_{\omega}\rightarrow 0$. Added to these puzzling findings is the
fact that the electronic specific heat continues to behave more or less in
a Fermi-liquid manner over the entire doping range from the overdoped
metal to the insulator\cite{yoshida,komiya,brugger,li}. These results
clearly demonstrate that a non-Fermi-liquid or `strange metal'
superconductor emerges from the Fermi-liquid background as doping is
reduced. Notably, there is considerable controversy over what constitutes
non-Fermi-liquid behavior and many proposals have been made to understand
its possible origin, ranging from preformed $d-$wave pairs\cite{senthil}
to fluctuating spin-density waves\cite{sachdev}, from the
Hubbard\cite{Vidhyadhiraja,sakai} and $t-J$\cite{eder} models to the
anti-de Sitter/conformal field theory (AdS-CFT)
correspondence\cite{zaanen}.

Insight into the origin of non-Fermi-liquid behavior comes from
recent angle-resolved photoemission spectroscopy
(ARPES)\cite{nparm}, Hall effect\cite{hallong}, and quantum
oscillation\cite{qoleyraud,qohelm,qovignolle} experiments, which
reveal the presence of FS reconstruction as a robust feature of
both electron and hole doped cuprates, suggesting that the ground
state involves some superlattice order. In this article, we
investigate a model of the underdoped cuprates with a density wave
ordered ground state and find that when the Fermi surface (FS)
breaks into pockets, the self-energy develops a strong momentum
dependence.  Analytic forms for various renormalization factors
are presented to delineate how the non-Fermi-liquid physics
arising from FS reconstructions can be understood at a
quantitative level. Our study thus provides a tangible model for
reconciling the seemingly contradictory doping evolutions of the
QP dispersion, the QP spectral weight and the specific heat in the
cuprates.

The possibility of FS reconstruction in the cuprates has been proposed
many times since it was found that the Hall density scales with
doping\cite{Ong} -- i.e., it is proportional to the small pocket area
rather than the large FS. We model this FS reconstruction by assuming a
spin density wave (SDW) order which results in a
nearly-antiferromagnetic-Fermi-liquid (NAFL) phase\cite{Pines}. In
addition to SDW order, we have analyzed other candidates for the competing
order including charge, flux, or $d-$density waves\cite{tanmoytwogap} and
find that the pseudogap symmetry which is the essential ingredient for the
origin of non-Fermi-liquid physics is insensitive to the particular nature
of the competing order state. We calculate the self-energy due to spin and
charge fluctuations within a QP-GW formalism described in Appendix A. The
model is known to properly describe the high-energy 'waterfall' features
in ARPES\cite{water3,susmita,watermore} as well as the doping evolution of
the optical spectra and the finite QP lifetime.\cite{tanmoyop} Within the
QP-GW model the self-energy splits the spectrum into coherent in-gap
states and an incoherent residue of the undoped upper and lower Hubbard
bands (U/LHBs). With underdoping, the {\it in-gap states} develop an SDW
gap and split into upper and lower magnetic bands
(U/LMBs)\cite{tanmoytwogap}, and the resulting `four-band' features are
consistent with quantum Monte Carlo (QMC) calculations\cite{grober}
[Appendix A]. The involvement of a quantum critical point (QCP) in the
optimal doping regime where the SDW order disappears is strongly
suggestive of an intermediate strength for correlations in the
cuprates.\cite{tanmoyop} Accordingly, there have been several recent
attempts to develop an intermediate coupling model for the cuprates
starting from either the weak-coupling limit, as in the present
calculation, or the strong-coupling limit\cite{paramekanti}. Therefore, we
compare our results with experiments as well as with the variational
cluster calculations of Paramekanti, Randeria and Trevedi
(PRT)\cite{paramekanti}, which approach the problem from the RVB limit.

This article is organized as follows.  In sections IIA and IIB, we
investigate the doping evolution of the spectral weight and that of
dispersion renormalization, respectively. The specific heat results are
presented in Section III. The discussion and conclusions are given in
Sections IV and V respectively. A summary of the underlying QP-GW model is
provided in Appendix A.  Various renormalization factors invoked are
summarized in Appendix B for the reader's convenience. Appendices C and D
address details of analytic forms for the renormalization factors and
those of our specific heat evaluation.

\section{Nearly-antiferromagnetic-Fermi-liquid renormaliztion factors}

\subsection{Momentum density and spectral weight renormalization}

We evaluate the exact values of the renormalization factors from FS
discontinuities of spectral function moments $M_{l}({\bf k})$ of various
order $l$,
\begin{eqnarray}\label{moments} 
\int_{-\infty}^{\infty}d\omega\omega^{l}A({\bf k},\omega) f(\omega)
\end{eqnarray}
These moments provide important information about the
spectral weight distribution in energy and momentum space as a function of
doping.\cite{paramekanti} The spectral density $A({\bf k},\omega)$
involves both the coherent QP and the incoherent part. Due to the Fermi
function $f(\omega)$, the moments $M_l$ display singularities at the Fermi
momentum $k_F$ which are characteristic of coherent {\it gapless}
quasiparticle excitations.  We consider first the zeroth order moment
which is simply the momentum density $n({\bf
k})$.\cite{foot3,Tanaka,ft_posi1} The interacting FS is determined from
the jump in $n({\bf k})$ at $k_F$ of the quasiparticles. The magnitude of
this jump defines the spectral weight renormalization $Z_{\omega}$. The
incoherent part in the spectral weight substantially modifies the shape of
$n({\bf k})$.
\begin{figure}[top]
\rotatebox{270}{\scalebox{0.5}{\includegraphics{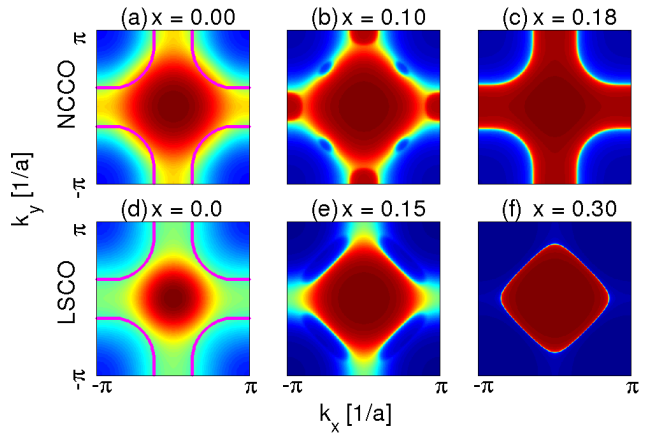}}}
\caption{(Color online) Momentum
density, $n({\bf k})$ as calculated using Eq.~\ref{moments} is shown
for various dopings for NCCO in (a-c) and for LSCO in (d-f).} \label{md2d}
\end{figure}

Fig.~\ref{md2d} shows maps of $n({\bf k})$ throughout the first
Brillouin zone as a function of doping for Nd$_{2-x}$Ce$_x$CuO$_4$
(NCCO) and hole doped La$_{2-x}$Sr$_x$CuO$_4$ (LSCO). In the
present NAFL case the combination of self-energy and SDW coherence
factors leads to characteristic structures in $n(k)$ at all
dopings {\it including half-filling}. At half-filling $n({\bf k})$
shows a maximum at the $\Gamma-$point and away from that it
decreases gradually and smoothly, from inside to outside the
LDA-like FS [magenta solid line in Fig.~\ref{md2d}(a) and (d)]. As
we dope the system with electrons, the spectral weight increases
at the $\Gamma-$point and in addition, $(\pi,0)$ and its
equivalent points largely gain spectral weight due to the
development of electron pockets in NCCO [Fig.~\ref{md2d}(b)]. With
further increase of doping, the FS undergoes two topological
transitions\cite{tanmoyprl,tanmoysns} as also reflected in the
momentum density calculations here. The first topological
transition in $n({\bf k})$ occurs when the LMB approaches the
Fermi level ($E_F$) and forms hole-like pockets at
$(\pm\pi/2,\pm\pi/2)$. For $x=0.15$, the hole pockets are fully
formed and they as well as the electron pockets increase in size
with further doping.  At $x=0.18$ the electron and hole pockets
merge at the hot-spot, the SDW gap collapses, and the full
metal-like $n({\bf k})$ appears [second topological transition].
For hole doping, the FS topological transition is complimentary to
the electron doped one and here the hole pocket appears first as
shown in Fig.~\ref{md2d}(e) and above the QCP, the electron-like
full FS appears in Fig.~\ref{md2d}(f).

\begin{figure*}[top]
\rotatebox{270}{\scalebox{0.7}{\includegraphics{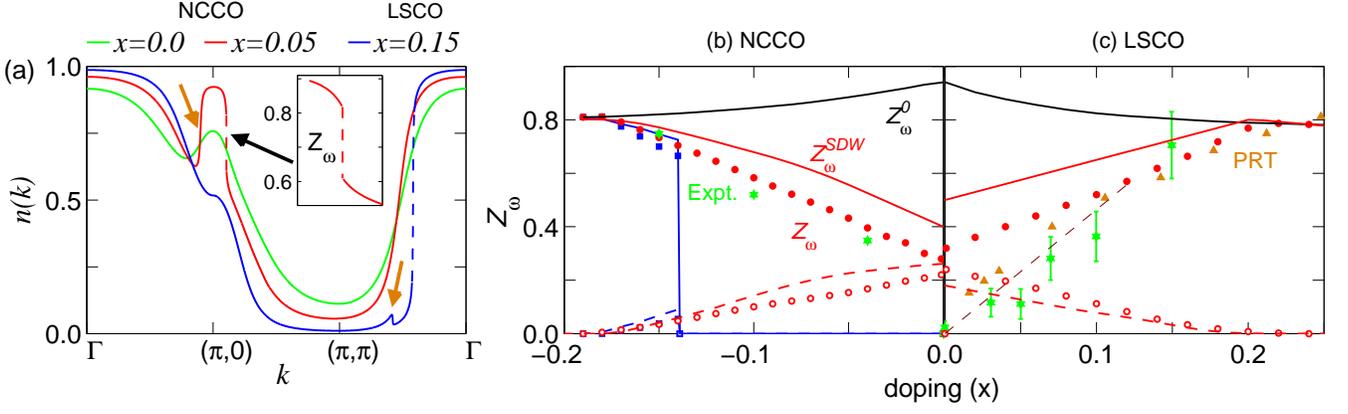}}}
\caption{(color online) (a) $n({\bf k})$ for NCCO and LSCO at
several representative dopings with dashed lines showing the
discontinuous jump at $k_F$ (highlighted in {\it inset}). Gold
arrows mark features from the shadow bands which cross $E_F$. (b)
$Z_{\omega}$ at $E_F$ are shown along the antinodal (red) and
nodal (blue) direction for NCCO. Filled (open) symbols give the
main (shadow) bands, compared with their corresponding analytical
approximation $Z_{\omega}^{SDW}$ plotted by solid (dashed) line of
same color. ARPES result (green) is extracted from
Ref.~\onlinecite{nparm}. (c) Same as (b) but for LSCO along the
nodal direction. These results are compared with PRT's
calculations for hole doping\cite{paramekanti} and ARPES results
\cite{yoshida} for LSCO along the nodal direction. All the
experimental data and PRT data in (b) and (c) are normalized to
highlight their doping evolution. Brown dashed line shows that if
there is nanoscale phase separation in LSCO then $Z_{\omega}$
would scale linearly with doping in the extreme underdoped
region.} \label{md1d}
\end{figure*}

A more quantitative account of the effect of self-energy corrections
on the residual coherent QP spectral weight
is provided in Fig.~\ref{md1d}(a), which shows $n({\bf k})$ along
high-symmetry lines for NCCO as well as LSCO.
Some important effects of correlations on the insulating state should
be noted here. At $x=0$,
$n[{\bf k}=\Gamma]\approx0.9$ and $n[{\bf k}=(\pi,\pi)]\approx0.1$,
implying that the self-energy redistributes
the spectral weight from the filled states to the unfilled regions
even in the insulating phase. PRT [with a
different parameter set $t=300$~meV, $t^{\prime}=-t/4$ and $U=12t$]
find a similar result $n[{\bf k}=\Gamma]
\approx0.85$\cite{paramekanti}. At half-filling $n({\bf k})$ is a
smoothly varying function throughout the BZ,
due to the absence of gapless quasiparticle for both NCCO (green line) and LSCO
(not shown). As we increase electron
doping, the spectral weight at $\Gamma$ [$(\pi,\pi)$] gradually
increases [decreases] whereas the spectral
weight increases rapidly at $(\pi,0)$, due to the development of
electron pockets, while discontinuities in
$n({\bf k})$ arise at the Fermi surface. In the underdoped region,
$n({\bf k})$ shows additional singularities
along $\Gamma\rightarrow (\pi,0))$ and
$(\pi/2,\pi/2)\rightarrow(\pi,\pi)$ due to the presence of shadow
bands as marked by gold arrows.
Since the shadow bands are usually weak in the
cuprates\cite{shadow,chang,he,yang1,meng},
experimental data are available predominantly along the arcs -- that is,
along the antinodal direction for NCCO and nodal
direction for LSCO.  These are compared with our theory in Fig.~\ref{md1d}(b),
effects of the ARPES matrix element notwithstanding.\cite{Sahra}

We define a coherent spectral weight renormalization factor from the
discontinuities in $n(k)$:
\begin{equation}\label{Zw}
Z_{\omega}=\Delta n(k_F),
\end{equation}
plotted as a function of doping for NCCO in
Fig.~\ref{md1d}(b)\cite{paramekanti}.
For the main band in NCCO along the antinodal direction, the UMB
crosses the Fermi level at all finite dopings
(electron pocket) and $Z_{\omega}$ decreases smoothly with underdoping
but vanishes discontinuously at $x=0$.
Shown also in Fig.~\ref{md1d}(b) is the conventional paramagnetic Green's
function renormalization factor
\begin{equation}\label{Zw0}
Z_{\omega}^0 = \left(1-\frac{\partial\Sigma^{\prime}(\omega)}
{\partial\omega}\right)_{\omega=0}^{-1}.
\end{equation}
It can be seen that the doping dependence in $Z_{\omega}^0$ is very
weak and strikingly opposite to
$Z_{\omega}$. Thus the strong doping dependence of $Z_{\omega}$ is
governed by the SDW gap collapse. In Appendix~C we derive
an approximate analytical form for the SDW corrected renormalization
factor $Z_{\omega}^{SDW}$ (at $\omega=0$) in terms of the
SDW coherence factors [see Eq.~\ref{A2Zw}]
\begin{equation}\label{ZwAF}
Z_{\omega}^{SDW}= \frac{Z_{\omega}^0}{2}\left[1\pm\left(1
+\left(\frac{2\Delta}{\xi_{k_F}-\xi_{k_F+Q}}\right)^2\right)^{-1/2}
\right].
\end{equation}
Note however that $k_F$ is the true Fermi momentum in the SDW
state, not the LDA one.  We see in Fig~\ref{md1d}(b) that this
simple analytic form captures the essential doping dependence of
the full $Z_{\omega}$. As doping increases, the weight of the
shadow bands (above the magnetic Brillouin zone) decreases [open
symbols in Fig.~\ref{md1d}(b)], with the corresponding spectral
weight shifting to the main bands. Along the nodal direction,
$Z_{\omega}$ remains zero up to optimal doping, and then shows a
jump to a slowly increasing value as the LMB crosses the Fermi
level. These results are in excellent agreement with
experiment\cite{nparm}.

A similar doping dependence of
$Z_{\omega}$ is observed in LSCO along the nodal direction, including the
jump at $x=0$ and the spectral weight transfer from the shadow band to the
main band\cite{shadow} in Fig.~\ref{md1d}(b).  This is in qualitative
agreement
with the calculation of PRT\cite{paramekanti} (triangles in
Fig.~\ref{md1d}(b)),
and with ARPES
measurements on LSCO\cite{yoshida} (open circles) above the optimal doping
region. However, in the very underdoped region, the ARPES data seem to
extrapolate smoothly to zero at half-filling (dashed line), which may be
related to nanoscale phase separations believed to be significant in LSCO.

Note that the analytic renormalization factor $Z_{\omega}^{SDW}$
is defined in the SDW phase, but is treated as a renormalization of the
paramagnetic phase dispersion.  This subtle point is an attempt to treat
the pseudogap physics, and
is discussed in more detail in Section IV below.

\subsection{FERMI VELOCITY AND DISPERSION RENORMALIZATION}

\begin{figure*}[top]
\rotatebox{270}{\scalebox{0.7}{\includegraphics{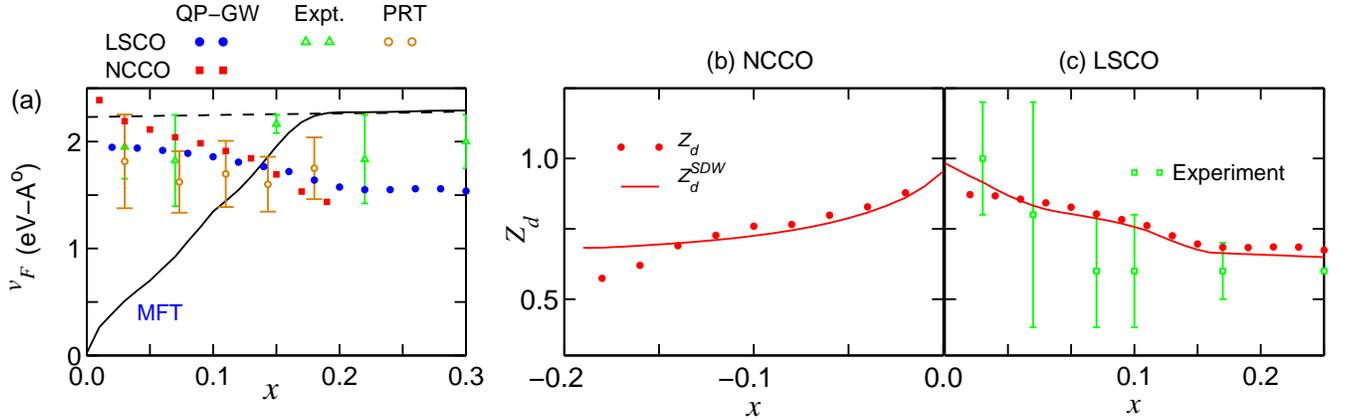}}}
\caption{(color online) (a) Fermi velocity $v_F$ along the
antinodal direction in NCCO and nodal direction in LSCO is
compared with ARPES data on LSCO\cite{yoshida,yang} and with PRT's
calculation\cite{paramekanti}. The blue dashed (solid) line gives
the corresponding LDA (MFT) results for LSCO. (b) Dispersion
renormalization $Z_d$ for NCCO along antinodal direction, compared
with an approximate analytical formula for the dispersion
renormalization $Z_d^{SDW}$ (solid line). (c) Same as (b) but
for LSCO along the nodal direction. The results are compared with
experimental data\cite{sahrakorpi}.} \label{M1}
\end{figure*}

We turn next to the first order spectral moment $M_1({\bf k})$.
One can measure the dispersion renormalization
from the size of the slope discontinuity, which
can be written as (Eq.~\ref{A2dM1})
\begin{equation}\label{dM1}
\Delta\left(dM_1({\bf k})/dk\right)_{k_F}= Z_{\omega}v_F,
\end{equation}
where $v_F$ is the Fermi velocity\cite{paramekanti}. Knowing
$Z_{\omega}$ from Figs.~\ref{md1d}(b) and~(c), we can
extract $v_F$ as a function of doping, as seen in Fig.~\ref{M1}(a). The
results
are obtained for both LSCO and NCCO
and compared with ARPES data on LSCO\cite{yoshida} and with PRT's
variational cluster
calculations\cite{paramekanti}(hole doping) as well as with LDA and
mean-field theory for LSCO. Mean-field results for NCCO
have an equivalent doping dependence and thus are not shown here.
Notably, in
spite of the substantial decrease of $Z_{\omega}$, the QP velocity
$v_F$ does not diminish on entering into the
pseudogap region. Although in the SDW mean field case $v_F$ decreases
smoothly with underdoping as the gap grows,
when a self-energy is introduced, the reduction of the coherent
spectral weight ($Z_{\omega}$) with underdoping
compensates for this, leading to a net enhancement of $v_F$. The
results are consistent with PRT and ARPES.
Similar results are also obtained in a self-consistent Born
approximation with antiferromagnetic pseudogap in
the $t-J$ model\cite{Prelovsek}.

We define a dispersion renormalization factor in terms of the
conventional band velocity with respect to its bare (LDA)
counterpart $(v_F^0)$ as
\begin{eqnarray}\label{Zd}
Z_d=v_F/v_F^0,
\end{eqnarray}
plotted in Figs.~\ref{M1}(a).
 In the paramagnetic case, dispersion renormalization is defined
as
\begin{eqnarray}\label{Zd0}
Z_d^0=Z_{\omega}^0Z_{k_F}^0=Z_{\omega}^0\left(1+\partial\Sigma'/v_F^0
\partial k\right).
\end{eqnarray}
This implies that in a conventional picture the deviation in
behavior of $Z_d^0$ from spectral weight renormalization
$Z_{\omega}^0$ comes solely from the $k-$dependence of the
self-energy. Since this self energy is approximately
$k$-independent, one finds $Z_d^0\approx Z_{\omega}^0$. In sharp
contrast, the calculated $Z_{d}$ [Fig.~\ref{M1}(b)] shows a
strikingly opposite doping dependence to $Z_{\omega}$. This can be
understood to be the result of an SDW gap, which introduces a new
$k$-dependence in the
dispersion renormalization as given by (Eq.~\ref{A2ZdAF}):
\begin{equation}\label{ZdAF}
Z_d^{SDW}=Z_{\omega}^0Z_{{\bf k}_F}^{SDW}=Z_{\omega}^0
\left(1+\frac{\Delta^2}{\xi_{k_F}\xi_{k_F+{\bf Q}}} \right).
\end{equation}
Figs.~\ref{M1}(b) and (c) compare $Z_{d}$ and $Z_d^{SDW}$ for NCCO and
LSCO respectively.
Thus the doping dependence of $Z_{d}$ implies that as we go towards
the Mott insulator, the dispersion tends towards the LDA-bands,
consistent with LSCO results (blue open circles)\cite{sahrakorpi}.
The opposite doping dependences of $Z_d$ and $Z_{\omega}$ can be
readily understood by comparing the corresponding analytical
formulas $Z_d^{SDW}$ (Eq.~\ref{ZdAF}) and $Z_{\omega}^{SDW}$
(Eq.~\ref{ZwAF}). $Z_d^{SDW}~(Z_{\omega}^{SDW})$ varies with SDW
gap as $\Delta~(1/\Delta)$, increasing (decreasing) with
underdoping. This in turn moves the in-gap states further away
from the Fermi level (hence decreasing $Z_{\omega}$), and thus
shifting the band towards the LDA values (increasing
$Z_{d}$)\cite{tanmoysw}.

\section{Specific heat }
\begin{figure}[htop]
\rotatebox{270}{\scalebox{0.5}{\includegraphics{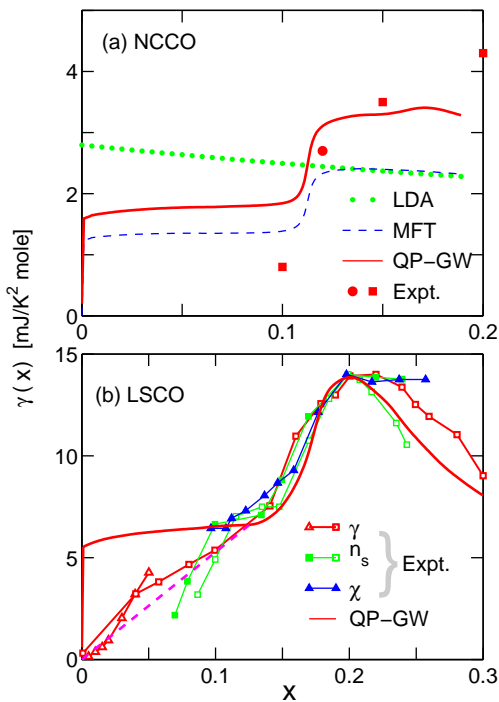}}}
\caption{(color online) (a) Specific heat
coefficient $\gamma(x)$ for different theoretical calculations
(various lines), and the experimental results on
NCCO (red squares)\cite{brugger} and PLCCO
(red circles)\cite{li}. (b) Same as (a) but for LSCO (red
squares\cite{yoshida} and triangles\cite{komiya}). Results are compared
with
$n_s$ [filled green squares\cite{yoshida} and open green
squares\cite{yang}] and $\chi$
(blue)\cite{ino}), all normalized at the VHS. The red
dashed line shows that in underdoped LSCO $\gamma$ scales linearly with
doping. Theoretical $\gamma$ has been scaled by a factor of 1.1,
consistent with a weak electron-phonon renormalization.}\label{fspheat}
\end{figure}

A striking application of the renormalization effects can be
seen in the doping evolution of the specific
heat. A general expression for the specific heat $c_V$ in the SDW state
including the self-energy correction is
given in Appendix D. We find that the Sommerfeld coefficient $\gamma$ can
be well
described in terms of the density of state (DOS) ($N(0)$) in the SDW state
as\cite{abrikosov},
\begin{eqnarray}\label{gamma}
\gamma = c_V(T)/T \approx \frac{2\pi^2k_B^2}{3}N(0)/Z_{\omega}^0,
\end{eqnarray}
where $N_0(0)$ is the mean-field DOS with SDW gap but without self energy
corrections.

Fig.~\ref{fspheat} compares experimental values of $\gamma$ as a function of doping in
LSCO\cite{yoshida,komiya} and electron doped NCCO\cite{brugger} and
Pr$_{1-x}$LaCe$_x$CuO$_4$ (PLCCO)\cite{li} with several calculations
including bare LDA, MFT results with SDW gap, and with self-energy
correction for both NCCO and LSCO. The striking differences between NCCO
and LSCO away from half-filling are due to the presence of the van-Hove
singularity (VHS) near $E_F$ in the latter case. $N(0)$ in LDA thus
decreases (from a finite value at $x=0$) with increasing electron doping,
whereas for LSCO $N(0)$ has a peak at the doping corresponding to the VHS
$x_{VHS}\sim 0.20$\cite{sahrakorpi}. SDW order opens a gap, reducing
$N(0)$, and introduces steps associated with the collapse of the SDW gap.
Thus, for electron doping $N(0)$ is nearly flat for $x<0.11$, reflecting
the quasi-two dimensionality of the electron pocket (constant DOS) in
cuprates. The step at $x\sim 0.11$ signals the appearance of the hole
pocket. Similarly in LSCO the appearance of the electron pocket near $(\pi
,0)$ at $x\approx 0.17$ within our model greatly enhances the VHS
features. Finally, adding self-energy corrections (in the form of
$1/Z_{\omega}^0$) preserves the general shape of the SDW $\gamma$, while
shifting its magnitude back towards the LDA values. It is interesting to
relate the doping dependence of $\gamma(0)$ and $Z_{\omega}$ in
Fig.~\ref{md1d}(b).
Note that $Z_{\omega}$ is evaluated at a particular Fermi momentum whereas
$\gamma(0)$ is computed after summing over $Z_{\omega}$ at all $k_F$. As
$Z_{\omega}$ exhibits complimentary doping dependencies for main and
shadow bands, so the total remains fairly constant.  These results are in
striking contrast to the strong coupling limit where $\gamma$ should
diverge with the effective mass as $x\rightarrow 0$.\cite{KOg}

The agreement with experiment is quite good in LSCO for $x\ge 0.10$,
including the strong VHS feature. Interestingly, the superfluid
density $n_s$ and the paramagnetic susceptibility data, which are
proportional to the total FS areas, show a similar doping dependence to
$\gamma$. However, below $x=0.10$ $\gamma\rightarrow 0$ as $x
\rightarrow 0$, an effect not captured by our calculations. Such an
effect could be due to a Coulomb gap\cite{komiya} and/or nanoscale
phase separation. The linear dashed line in Fig.~3(b) illustrates
the corrected form expected in the latter case.  Note that nanoscale
phase separation would produce the dashed line seen in
Fig.~\ref{md1d}(c) as well as explain the anomalous doping dependence of
the
chemical potential.\cite{FIMT}  Hence the enhanced gossamer features seen
in LSCO may be related to nanoscale phase separation.

\section{discussion}

The present calculations are most appropriate for electron doping,
where only the $(\pi ,\pi )$ commensurate SDW
order is observed, and the model is in very good agreement with
experiment. Remarkably, the same model when applied
to hole-doped cuprates describes many aspects of the two-gap
scenario\cite{tanmoytwogap,AndHir}, despite the
fact that it does not capture the incommensurate magnetization.
The non-Fermi-liquid behavior presented here is not
sensitive to the specifics of the competing order, but only to the
resulting superlattice $q$-vector.  We have analyzed other candidates for
the competing order including charge, flux, or $d-$density
waves\cite{tanmoytwogap},
and find that the results are insensitive to the
nature of the competing order state.  Thus, Eqs. 4 and 8 continue to hold
for any $Q=(\pi ,\pi )$ order, as long as the appropriate gap $\Delta$ is
used.

In the presence of long-range magnetic order, the Green's function
develops a second pole, and should properly be treated as a tensor.
However, the pseudogap phase is more likely to be associated with only
short range order, in which case the second pole does not cross the real
axis, the Green's function can be treated as a scalar, and {\it the
information about incipient gap formation is encoded in the
self-energy}.\cite{MKII} The analytic approximations of Eqs. 4 and 8
mimic
this effect by treating the renormalization factors as acting on the
paramagnetic dispersion $\xi_{{\bf k}}$, with information on proximity to
long-range magnetic order encoded into $Z_{\omega}^{SDW}$ and
$Z_{d}^{SDW}$.

This approach is similar to the phenomenological model introduced by Yang
et al.\cite{YRZ} to describe RVB physics.  Indeed, their phenomenological
self energy is quite similar to the form expected for a NAFL, except for
the treatment of scattering at the magnetic zone boundary, which splits
the pockets expected for AF order into two half-pockets. One puzzle is
that the effect of strong scattering at a superlattice zone boundary is
already well understood in standard Lifshitz-Kosevich theory\cite{magb},
where it leads to a quantum switching between arcs of Fermi surface.
This produces harmonic mixing frequencies in a quantum oscillation (QO)
spectrum, and it is hard to see how it could evolve into the
short-circuiting effect of Yang, et al.  At any rate, the QOs observed in
NCCO are more consistent with the present model.\cite{qohelm}

\section{conclusion}

In conclusion, we have shown that a number of salient features of the
non-Fermi-liquid state of the underdoped cuprates can be understood within
the framework of a competing density wave order, which breaks the
particle-hole symmetry, and drives reconstruction of the FS. We provide a
transparent and analytic basis for describing how the non-Fermi-liquid
effects play out in renormalizing spectral weight via $Z_{\omega}$ and
electronic dispersion via $Z_d$, and how they conspire to yield a specific
heat in the cuprates which is essentially conventional in nature at all
dopings. Our framework would provide a straightforward basis for
understanding how the broken-symmetry order leads to `non-Fermi-liquid'
effects,
not only in the cuprates, but also in heavy-fermions\cite{aronson},
Fe-based
superconductors\cite{xu} and other strongly correlated materials.

\begin{acknowledgments}
This work is
supported by the U.S.D.O.E grant DE-FG02-07ER46352, and benefited
from the allocation of supercomputer time at NERSC and
Northeastern University's Advanced Scientific Computation Center
(ASCC).
\end{acknowledgments}

\appendix
\addcontentsline{toc}{section}{Appendix}
\section{Details of quasiparticle-GW (QP-GW) Model}

\begin{figure}[top]
\rotatebox{270}{\scalebox{0.35}{\includegraphics{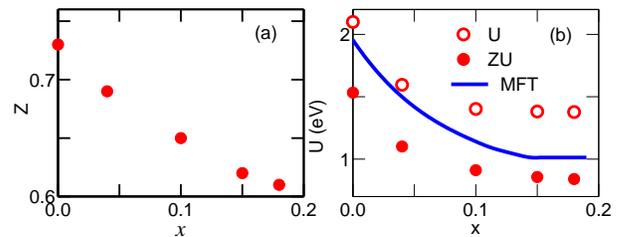}}
} \caption{(color online) (a) Average renormalization factor $Z$
decreases linearly with doping, very much like $Z_{\omega}^0$ in
Figs.~\ref{md1d}(b) and (c). (b) Our computed doping dependence of selfconsistent
values of $U$
and $ZU$ is compared with earlier mean field
results\cite{kusko,tanmoyprl}.}\label{UZ}
\end{figure}

\begin{figure}[top]
\rotatebox{270}{\scalebox{0.35}{\includegraphics{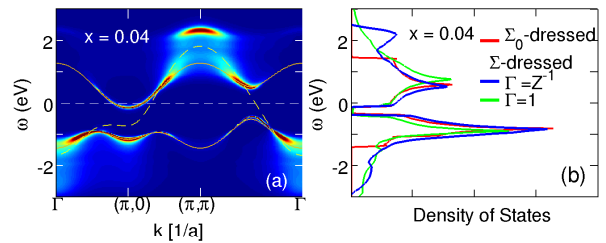}}
} \caption{(color online) (a) Spectral intensity as a function of
$\omega$ along the high-symmetry lines for several dopings (at
temperature $T = 0$). Blue to red color map gives the minimum to
maximum intensity. The yellow dashed line gives the underlying LDA
dispersion where the gold lines represent the renormalized
magnetic bands ($\Sigma_0$-dressed). (b) The QP-GW DOS (blue
lines) is compared with $\Sigma_0$-dressed DOS (red line)
calculated at $T=0$. The green lines show the DOS without the
vertex correction. }\label{selfcon}
\end{figure}
In the QP-GW formalism the bare dispersion is taken
as the LDA dispersion ($\xi_{{\bf k}}=\epsilon_{{\bf k}}-E_F$),
modeled via a tight-binding (TB)
fit\cite{markietb,foot2,ft_sophi1,ft_sophi3,ft_sophi4}. We
calculate the self-energy in a GW-like formalism using a simplified (one
parameter) scheme where the input and
final self-energies are self-consistent in the coherent part
only (QP-GW model).

In the following, we use a `tilde' over a quantity to
symbolize that it is a $2\times2$ matrix. The
self-energy in the underdoped region is written in canonical form
using the Nambu formalism as
\begin{eqnarray}\label{Asigma}
\tilde\Sigma^{t}({\bf k},\sigma,i\omega_n)= \tilde\Sigma({\bf
k},\sigma,i\omega_n) + \phi({\bf
k},\sigma,i\omega_n)\tilde{\tau_1}.
\end{eqnarray}
The dressed Green's function is then
\begin{eqnarray}\label{AGdressed}
\tilde G^{-1}({\bf k},\sigma,i\omega_n) = i\omega_n\tilde{1}
-\tilde H_{LDA}-\tilde\Sigma^t({\bf k},\sigma,i\omega_n)
\end{eqnarray}
where $\tilde H_{LDA}$ is the bare Hamiltonian defined in the
magnetic zone as $diag[\xi_{{\bf k}},~\xi_{{\bf k}+{\bf Q}}]$, and
the $\tilde\tau_i$ are Pauli matrices. At each step, we adjust the
chemical potential $E_F$ to fix the doping.  Real frequency Green's
functions are extracted from the Matsubara results by analytic
continuation $i\omega_n \rightarrow \omega+i\delta$. Here, $\tilde\Sigma
({\bf k},\sigma,i\omega_n)$ is a
$2\times2$ matrix in the SDW state\cite{SWZ}, whose diagonal part
renormalizes $\tilde H_{LDA}$, while the off-diagonal term gives
a (small) anomalous frequency dependence to the gap function
$\phi({\bf k},\sigma,i\omega_n)=\sigma\Delta$.  $\Delta=US$ is the
magnetic gap parameter and
$S$ is the magnetization at the commensurate vector ${\bf
Q}=(\pi,\pi)$ for Hubbard $U$, which is calculated using a
mean-field approximation\cite{tanmoytwogap}. The doping dependence
of the on-site Hubbard $U$ is obtained due to charge screening from
$U=\left<V(q)/(1+V(q)\chi(q)\right>$, where $V(q)$ is the
long-range Coulomb interaction\cite{markiecharge}, and $\chi(q)$
is the charge susceptibility in the $\phi-$gapped state defined
below. The obtained values of screened $U$ are given in
Ref.~\onlinecite{tanmoyop} and plotted in Fig.~\ref{UZ}(b).
\\
\\
We calculate the self energy due to the spin as well as the charge
response within a GW framework as
\begin{eqnarray}\label{Aselfeng}
&&\tilde\Sigma({\bf k},\sigma,i\omega_n)=\frac{3}{2}U^2Z
\sum_{{\bf q},\sigma^{\prime}}^ {\prime}
\int_{-\infty}^{\infty}\frac{d\omega_p}{2\pi}\nonumber\\
&&\tilde G({\bf k}+{\bf q},\sigma^{\prime},i\omega_n+\omega_p)
\Gamma({\bf k},{\bf q},i\omega_n,\omega_p){\rm Im}[\tilde\chi_{\rm
RPA}^{\sigma\sigma^{\prime}}({\bf q},\omega_p)],\nonumber\\
\end{eqnarray}
where the prime over the momentum summation indicates that the summation
is restricted within the magnetic Brillouin zone. The dressed
susceptibility in the above equation is given
in terms of the $2\times 2$ RPA susceptibility as,\cite{SWZ}
\begin{equation}\label{ARPA}
\tilde\chi_{\rm{RPA}}^{\sigma\sigma/\sigma\bar{\sigma}}({\bf
q},i\omega_
 n)
=\frac{\tilde\chi_0^{\sigma\sigma/\sigma\bar{\sigma}}({\bf
q},i\omega_n) } {\tilde{{\bf 1}}\pm
U\tilde\chi_0^{\sigma\sigma/\sigma\bar{\sigma}}({\bf
q},i\omega_n)}.
\end{equation}
Here the superscript $(\sigma\sigma)$ refers to the combined
charge plus longitudinal spin susceptibility tensor,
whereas $(\sigma\bar{\sigma})$
(with $\bar{\sigma}=-\sigma$) gives the transverse susceptibility
tensor, and the
$\tilde\chi_0^{\sigma\sigma/\sigma\bar{\sigma}}({\bf
q},i\omega_n)$ are bare susceptibilities.
\newline
\\
\\
A self-consistent `dressed' GW
calculation would include $\Sigma$ on the right-hand side of
Eq.~\ref{Aselfeng} in both $G$ and $\chi_0$.  This
generally leads to problems unless a vertex correction $\Gamma$ is
included. Improved results are
often found by using a `bare' $G_0$ and $\chi_0$ -- bare in the sense of
not
including $\Sigma$. This latter $G_0W_0-$scheme also fails in the present
calculation
by producing too large a renormalization of the band dispersion.   This is
because the imaginary part of the bare susceptibility
has the form $\chi_0^{\prime\prime}\sim\delta
(\omega-[\xi_{k+q}-\xi_{k}])$, so that near the Fermi surface,
$\chi_0^{\prime\prime}$ should scale in frequency with the dressed
quasiparticle dispersion.  Since $G_0W_0$ uses the bare dispersions,
peaks in $\chi_0^{\prime\prime}$, which control the renormalization,
lie at too high an energy.

We therefore introduce a modified, or `quasiparticle' (QP) GW
approximation\cite{water3} for the right-hand side (RHS) of
Eq.~\ref{Aselfeng},
as follows.  In Eq.~\ref{Aselfeng}, we dress both $G_0$ and $\chi_0$ with
an
`input' self-energy chosen as
\begin{equation}\label{ASigmainput}
\tilde\Sigma(i\omega_n)=(1-1/Z)i\omega_n\tilde{1}.
\end{equation}
Thus, the input $\Sigma$ contains a single parameter $Z$, which gives
an overall renormalization of the `input'
dispersions (RHS of Eq.~\ref{Aselfeng}).  With this approximation, the
bare susceptibilities become
\begin{eqnarray}\label{AAFMChi}
\tilde\chi_0^{{\sigma\sigma}/{\sigma\bar{\sigma}}}({\bf
q},i\omega_n) =
-Z^2\sum_{{\bf k}}^{\prime}\sum_{\nu,\nu^{\prime}} \tilde
S_{\nu,\nu^{\prime}}^{\sigma\sigma/\sigma\bar{\sigma}}
\frac{f(Z{E}^{\nu}_{{\bf k}}) -f(Z{E}^{\nu^{\prime}}_{{\bf k}+{\bf
q}})} {i\omega_n+Z{E}^{\nu}_{{\bf k}}-Z{E}^{\nu^{\prime}}_{{\bf
k}+{\bf  q}}}.
\nonumber\\
\end{eqnarray}
The prime over the summation has the same meaning as in
Eq.~\ref{Aselfeng}. In Eq.~\ref{AAFMChi}, the
$\nu$ summation is over the two magnetic bands UMB ($\nu=+$) and LMB
($\nu=-$):
\begin{equation}\label{AE}
E_{{\bf k}}^{\pm}=(\xi_{{\bf k}}^+ \pm E_{0{\bf k}})
\end{equation}
 with
$\xi_{{\bf k}}^{\pm}=(\xi_{{\bf k}}\pm\xi_{{\bf k}+{\bf Q}})/2$
and $E_{0{\bf k}} =\sqrt{(\xi_{{\bf k}}^-)^2+\Delta^2}$
(Ref.~\onlinecite{tanmoysw}), $f(E)=1/(1+\exp{(E/k_BT)})$ is the
Fermi function at temperature $T$, and $k_B$ is the Boltzmann
constant. The coherence factors $\tilde
S_{\nu,\nu^{\prime}}^{\sigma\sigma/\sigma\bar{\sigma}}$ give the
amplitude of the scattering of the quasiparticles with the charge
and magnon modes of the system respectively with components
\begin{eqnarray}\label{AChicoh}
\tilde S_{\nu,\nu^{\prime}}^{\sigma\sigma/\sigma\bar{\sigma}}(11)
&=&(\alpha_{{\bf k}}\alpha_{{\bf k}+{\bf q}} \pm
\nu\nu^{\prime}\beta_{{\bf k}}\beta_{{\bf k}+{\bf q}})^2,\nonumber\\
\tilde S_{\nu,\nu^{\prime}}^{\sigma\sigma/\sigma\bar{\sigma}}(12)
&=&-\nu(\alpha_{{\bf k}}\beta_{{\bf k}}
\pm\nu\nu^{\prime}\alpha_{{\bf k}+{\bf q}}\beta_{{\bf k}+{\bf
q}}).
\end{eqnarray}
Here
\begin{eqnarray}\label{AAFMChib}
\alpha_{{\bf k}} (\beta_{{\bf k}}) = \sqrt{\frac{1}{2}
\left(1\pm\frac{\xi_{{\bf k}}^-}{E_{0{\bf k}}}\right)}
\end{eqnarray}
respectively are the weights associated with the U/LMBs. The other
coherence factors in Eq.~\ref{AChicoh} can be derived using the
translational symmetry with respect to ${\bf q}$.

A limitation of the present scheme is also apparent from
Eq.~\ref{Aselfeng}. This self-energy is a good
approximation for the coherent, dressed bands, but does not extend to
the incoherent part of the spectrum,
thereby underestimating the incoherent spectral weight.  We have
empirically found that this can be partly
remedied by incorporating a vertex function $\Gamma$.  Consistent with
the QP approximation, the vertex
correction to the self-energy is obtained through Ward's identity as
\begin{eqnarray}\label{Avertex}
\Gamma({\bf k},{\bf
q},i\omega_n,\omega_p)&=&1-(\partial\Sigma^{\prime}/
\partial\omega)_{\omega=\omega_0}=1/Z.
\end{eqnarray}
Even though we have chosen perhaps the simplest form of the vertex
correction,
it has a large impact on the spectral weight transfer.
$\Gamma=1/Z$ eventually reduces the renormalization of the bare
susceptibility in Eq.~\ref{Aselfeng} and hence the spectral weight
is spread out more towards higher energies, enhancing the incoherent
spectral weight. An illustration of the importance of this vertex
correction is given in Fig.~\ref{selfcon}(b).

\begin{figure}[h]
\rotatebox{0}{\scalebox{0.5}{\includegraphics{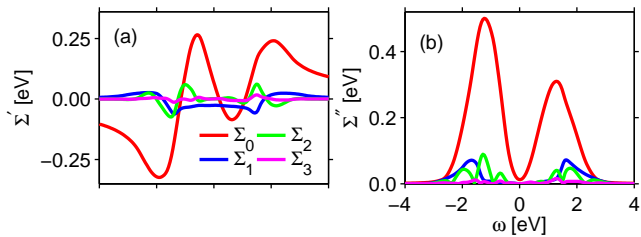}} }
\caption{(color online) Real and imaginary part of
the self-energy as expanded in the tight-binding form of
Eq.~A11.}\label{fse}
\end{figure}

Next we discuss our self-consistent scheme, and explain how the
parameter $Z$ is chosen.  We choose $Z$ to match the average
renormalization in the low-energy (coherent) part of the spectrum.
Specifically, if $\Sigma^{\prime}$ is the real part of the
diagonal self energy, then we adjust $Z$ self-consistently until
it satisfies $Z=(1-\partial\Sigma^{\prime}/\partial
\omega)^{-1}_{\omega=\omega_o}$,  where $\omega_o$ is an average
quasiparticle excitation energy, which is related to the poles in
$G$. This gives a good self-consistent result for the coherent
spectral weight in the low-energy region (see Fig.~\ref{selfcon}(a)),
whereas the incoherent parts in the higher energy regions are not
self-consistent. Thus our scheme is in the spirit of Landau's
quasiparticles, except that Landau assumed that all of the spectral
weight goes into the QP band, while we have only a fraction $Z$.

When this is done, we find that $U$ is effectively further
renormalized by the doping dependent $Z$ [Fig.~\ref{UZ}(a)],
so that the product $U\chi_0$ is approximately independent of $Z$.
The resulting $ZU$ closely resembles our earlier mean-field
calculations as shown in Fig.~\ref{UZ}(b).

Lastly, we find that the momentum dependence of the fluctuation
self-energy $\Sigma$ is
relatively weak\cite{water3,jarrell}. To emphasize
this point, we expanded the momentum dependence of the self energy in
a form similar to the tight-binding model:
\begin{eqnarray}\label{ASeTB}
\Sigma(\omega,{\bf k}) &=&
\Sigma_0(\omega)+\Sigma_1(\omega)\left(c_x+c_y\right)\nonumber\\
&&+\Sigma_2(\omega)c_xc_y+\Sigma_3(\omega)\left(c_{2x}+c_{2y}\right),
\end{eqnarray}
where $c_{\alpha (x/y)}=\cos{(\alpha k_{(x/y)}a)}$. We calculate
the self-energy at four high symmetry points ${\bf
k}=(0,0),~(\pi,0),~(\pi,\pi)~{\rm and}~(\pi/2,\pi/2)$ to obtain
the above coefficients as shown in Fig.~\ref{fse}. Clearly, only
the $k-$independent part (red line) has a strong contribution.
Therefore, we have simplified the self-energy calculation by
approximating it with a $k$-independent average value taken as
the value at $k=(\pi /2,\pi /2)$.  Since we are neglecting the
$k-$dependence of the self energy, $\Sigma_{11}=\Sigma_{22}$ and
$\Sigma_{12}=\Sigma_{21}$.

\begin{figure}[htop]
\rotatebox{270}{\scalebox{.6}{\includegraphics{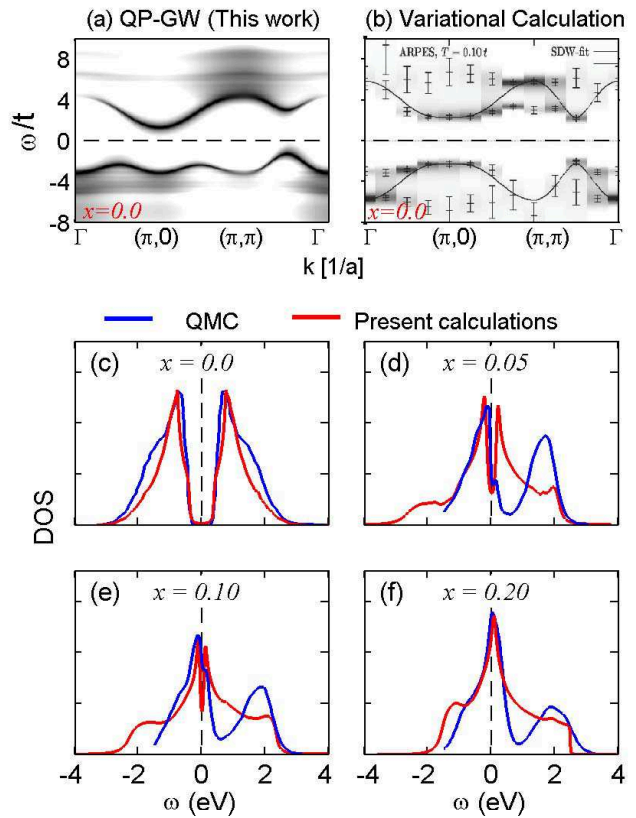}}}
\caption{(color online) (a) The spectral intensity of NCCO at
$x=0$, plotted (in logarithmic scale) along the high-symmetry
lines, is compared with variational calculations\cite{grober} in
(b). (c)-(f) The computed DOS at various dopings are compared with
the corresponding QMC results (blue lines) for x = 0.0
[Ref.~\onlinecite{maier}] and for x = 0.05 to x = 0.20
[Ref.~\onlinecite{jarrell}].}\label{swarpes}
\end{figure}

Figure~\ref{swarpes}(a) shows that in the QP-GW scheme the
spectral weight splits up into four band-like features.  The two
bands closer to the Fermi level are the UMB and LMB, associated
with the development of the spin-density wave (SDW). The residual
incoherent spectral dispersions at higher energies are the UHB and
LHB. Similar four band features are found in the variational
cluster calculations shown in Fig.~\ref{swarpes}(b)\cite{grober}.
The corresponding density of states (DOS) shows four peaks
associated with these four bands, Figs.~\ref{swarpes}(c)-(f). These
four band features separated by SDW gap or `waterfall' effects
show semi-quantitative agreement with QMC
results\cite{maier,jarrell}.  With doping, the two magnetic bands
merge in a QCP near optimal doping, while the two Hubbard bands
occur at the top and bottom of the LDA bands, and hence the
associated Hubbard band splitting is comparable to the LDA
bandwidth at all dopings.

\section{Summary of renormalization factors}
We summarize the various renormalization factors that arise in
this study for convenience reference. Equation numbers in square brackets
indicate where a specific factor is
discussed in the text.
\begin{eqnarray}\label{renorm}
Z_{\omega} &=& \Delta n(k_F)\qquad\qquad\qquad\qquad\qquad\qquad
[\text{Eq.~\ref{Zw}}]\nonumber\\
&=& {\rm Total~spectral~weight~renormalization}.\nonumber
\end{eqnarray}
\begin{eqnarray}
Z_{\omega}^0 &=& \left(1-\partial\Sigma^{\prime}/
\partial\omega\right)_{\omega=0}^{-1}\qquad\qquad\qquad\quad[\text{Eq.~\ref{Zw0}}
]\nonumber\\
&=&{\rm Self~energy~contribution~to~spectral~weight}\nonumber\\
&&{\rm renormalization}.\nonumber
\end{eqnarray}
\begin{eqnarray}
Z_{\omega}^{SDW}&=& \frac{Z_{\omega}^0}{2}\left[1\pm\left(1
+\left(\frac{2\Delta}{\xi_{k_F}-\xi_{k_F+Q}}\right)^2\right)^{-1/2}
\right]\nonumber\\
&&\qquad\qquad\qquad\qquad\qquad\qquad\qquad[\text{Eq.~\ref{ZwAF}}]\nonumber\\
&=&\text{Analytical formula for SDW spectral weight}\nonumber\\
&&{\rm renormalization}.\nonumber
\end{eqnarray}
\begin{eqnarray}
Z_{d}&=&v_F/v_F^0\qquad\qquad\qquad\qquad\qquad\qquad[\text{Eq.~\ref{Zd}}]
\nonumber\\
&=&{\rm Velocity~renormalization.}\nonumber
\end{eqnarray}
\begin{eqnarray}
Z_{d}^0&=&Z_{\omega}^0Z_{\bf k}^0
=Z_{\omega}^0\left(1+{\partial\Sigma^{\prime}}/
{v_{F}^0\partial k}\right)\qquad\quad[\text{Eq.~\ref{Zd0}}]\nonumber\\
&=&{\rm Conventional~dispersion~renormalization.}\nonumber
\end{eqnarray}
\begin{eqnarray}
Z_{d}^{SDW}&=&Z_{\omega}^0\left(
1+{\Delta^2}/{\xi_{k_F}\xi_{k_F+Q}}\right)
\qquad\qquad\qquad[\text{Eq.~\ref{ZdAF}}]\nonumber\\
&=&{\rm Analytical~formula~for~SDW~dispersion}\nonumber\\
&&{\rm renormalization.}\nonumber
\end{eqnarray}
\begin{eqnarray}
\gamma &=&
c_V/T\qquad\qquad\qquad\qquad\qquad\qquad\qquad[\text{Eq.~\ref{gamma}}]\nonumber\\
&=&{\rm Specific~heat~coefficient.}\nonumber
\end{eqnarray}

\section{Analytical form of various renormalization factors}
Near the Fermi level, the dressed Green's functions from
Eq.~\ref{AGdressed} can be approximated as,
\begin{eqnarray}\label{A2G11}
G_{11}({\bf k},\omega)&=&Z^0_{\omega}\frac{\omega-\bar{\xi}_{{\bf
k}+{\bf Q}}}
{(\omega-\bar{\xi}_{{\bf k}})(\omega-\bar{\xi}_{{\bf k}
+{\bf Q}})+(\overline{\Delta})^2}\nonumber\\
&=&\frac{Z^0_{\omega}}{\omega-Z_d^{SDW}({\bf k})\xi_{{\bf k}}}
\end{eqnarray}
where $\bar{\xi}_{{\bf k}}=Z^0_{\omega}\xi_{{\bf k}}$,
$\overline{\Delta}=Z_\omega^0\Delta$, and $\xi_{{\bf k}}$ is the
bare (LDA) dispersion calculated at the same doping. In
Eq.~\ref{A2G11}, $Z^0_{\omega}
=(1-\partial\Sigma_{11}^{\prime}(\omega)/\partial\omega)^{-1}$
(Eq.~\ref{Zw0}) is the part of the spectral weight renormalization
due to the self-energy, where we have neglected the contribution of
$\Sigma_{12}$. Even though we assume a $k-$independent self-energy
and thus $Z_{{\bf k}}^0=1$ from Eq.~\ref{A2G11}, the SDW gap
introduces a $k-$dependent dispersion renormalization (defined
here at $\omega=0$) given by
\begin{eqnarray}\label{A2ZdAF}
Z_d^{SDW} ({\bf k})= Z_{\omega}^0
\left(1 +\frac{\Delta^2}{\xi_{{\bf k}}\xi_{{\bf k}+{\bf
Q}}}\right).
\end{eqnarray}
Note that $\xi_{{\bf k}}$, $~\xi_{{\bf k}+{\bf Q}}$, and $\Delta$
are all bare values in Eq.~\ref{A2ZdAF} as the same
renormalization factor $Z_{\omega}^0$ gets cancelled out in the last
term. Furthermore, we can split the spectral function $A({\bf
k},\omega)={\rm Im}G({\bf k},\omega)/\pi$ into a coherent part (at
Fermi level) and an incoherent part where the former can be
represented by a delta function as
\begin{eqnarray}\label{A2A}
A_{11}({\bf k},\omega)&=& Z_{\omega}^0\left( \alpha_{{\bf
k}}^2\delta(\omega-\bar{E}^{+}_{{\bf k}}) +\beta_{{\bf
k}}^2\delta(\omega-\bar{E}^{-}_{{\bf k}})\right)_{\omega=0}
\nonumber\\
& +& A_{incoh}({\bf k},\omega>0)
\end{eqnarray}
where $\bar{E}_{{\bf k}}^{\pm}=Z_{\omega}^0E_{{\bf k}}^{\pm})$.
Therefore, the coherent part is governed by the SDW coherence
factors $\alpha_{{\bf k}}$ and $\beta_{{\bf k}}$ for the filled
state with a renormalization by $Z_{\omega}^0$. The zeroth and
first order moment of the spectral weight (Eq.~\ref{moments}) then
can be approximated respectively as\cite{paramekanti},
\begin{eqnarray}
n({\bf k}) &=& \int_{-\infty}^{0} A_{11}({\bf k},\omega)
d\omega\nonumber\\
&\approx&Z_{\omega}^0\left[\alpha_{{\bf k}}^2\theta(-\bar{E}_{{\bf
k}}^{ +})
+\beta_{{\bf k}}^2\theta(-\bar{E}_{{\bf k}}^{-})\right]\\
\label{A2nk}
M_{1}({\bf k}) &=& \int_{-\infty}^{0} A_{11}({\bf k},\omega)\omega
d\omega\nonumber\\
&\approx&Z_{\omega}^0({\bf k}-{\bf k}_F)\left[\alpha_{{\bf
k}}^2v_F^+ \theta(-\bar{E}_{{\bf k}}^{+})
+\beta_{{\bf k}}^2v_F^-\theta(-\bar{E}_{{\bf k}}^{-})\right]. \nonumber\\
\label{A2M1}
\end{eqnarray}
In the second equation above, we have expanded the renormalized
band near the Fermi level as $\bar{E}_{{\bf k}}^{\pm}\approx
v_{F}^{\pm}({\bf k}-{\bf k}_F)$, where $v_{F}$ is the
corresponding Fermi velocity. The singularity in $n({\bf k})$
becomes
\begin{eqnarray}
Z_{\omega}=\Delta n ({\bf k}_F) &=& Z_{\omega}^0\alpha_{{\bf
k}_F}^2\equiv Z_{\omega}^+
\qquad \text{for UMB}\nonumber\\
\qquad ~~~~~~&=& Z_{\omega}^0\beta_{{\bf k}_F}^2\equiv
Z_{\omega}^- \qquad \text{for LMB}.
\label{A2Zw}
\end{eqnarray}
Note that in the present case, the weight for UMB (LMB) appears
along the antinodal (nodal) direction when the band crosses the Fermi
level. Inserting the form of $\alpha,~\beta$ from
Eq.~\ref{AAFMChib}, we get Eq.~\ref{ZwAF}. Similarly, inserting
this $Z_{\omega}$ in Eq.~\ref{A2M1}, we can measure the singular
jump in $M_1({\bf k})$ as
\begin{eqnarray}\label{A2dM1}
\Delta \left(dM_{1}({\bf k})/d{\bf k}\right)&=&
Z_{\omega}^{\pm}v_F^{\pm}.
\end{eqnarray}
Thus $Z_{\omega}$ and $v_F$ acquire a $k$-dependence through the SDW
coherence factor.

\section{Specific Heat Calculation}

Following the derivation by Abrikosov {\it et. al.} \cite{abrikosov},
we calculate the entropy in the SDW state
for a strongly correlated system at finite temperature (in the low
temperature limit) as
\begin{eqnarray}\label{BS}
S(T) &=& -\frac{2\beta k_B^2}{2\pi
i}\int_{-\infty}^{\infty}d\omega\omega \left(-\frac{\partial
f(\omega)}{\partial
\omega}\right)\nonumber\\
&&~\times\sum_{{\bf k},\sigma}{\rm Tr}\left[\ln{\tilde
G_R^{-1}({\bf k},\sigma,\omega)}
-\ln{\tilde G_A^{-1}({\bf k},\sigma,\omega)} \right].\nonumber\\
\end{eqnarray}
Here the $2\times 2$ retarded and advanced (dressed) Green's functions
$G_R/G_A$ depend on the temperature
through the SDW order parameter only, which has a very weak dependence
in the low temperature region.  We can
rewrite Eq.~\ref{BS} in terms of a dimensionless parameter $y=\beta
\omega$ following Ref.~\onlinecite{toschi},
and then taking the temperature derivative we get the expression for
the specific heat as
\begin{eqnarray}\label{BcV}
c_V(T) = -\frac{k_B\beta^2}{4\pi}\int_{-\infty}^{\infty}dy y^2
~{\rm{sech}}^2(y/2)~~~~~~~~~~~~~~~~~~~~~~\nonumber\\
~\times\sum_{{\bf k},\sigma}{\rm Tr}\left[{\rm Im}\left(\tilde
G_R({\bf k},\sigma,\omega)\frac{\partial}{\partial \omega}\tilde
G_R^{-1}({\bf k},\sigma,\omega)\right)\right]_{\omega=y/\beta}.
\nonumber\\
\end{eqnarray}
$c_V (T)$ behaves linearly with $T$ in the low temperature region,
while the slope ($\gamma$) undergoes an
abrupt change with a kink in the waterfall region. Since the high
energy kink
energies for cuprates are around 0.3 to 0.6
eV, the kink should appear in $c_V$ only at a very high temperature,
$T_k\sim10^3 K$.  Thus, in our calculation
of $\gamma$ (Fig.~5), we have used the full expression for $c_V$ above
but evaluated it in the $\omega=0$ limit.

In the $T=0$ ($\omega\rightarrow0$) limit, Eq.~\ref{BcV}
can be simplified as $\rm{sech}(y)=\delta(y-\beta\omega)$.
In this limit, the
imaginary part of the self-energy is zero and thus the last quantity
in Eq.~\ref{BcV} is calculated by noting that the
$\omega-$derivative of the real part of the self-energy is:
diagonal term~= $1-1/Z_{\omega}^0$, and
off-diagonal term~=$(\partial \Sigma^{\prime}_{12}
/\partial\omega)_{\omega=0}$. This simplifies Eq.~\ref{BcV} as
\begin{eqnarray}\label{BcVA}
c_V&\approx &\frac{2}{3}\pi^2k_B^2T\sum_{{\bf k}}\left[
\frac{1}{Z_{\omega}^0}\left(A_{11}({\bf k},0)+A_{22}({\bf
k},0)\right)\right.\nonumber\\
&&\left.+\frac{\partial\Sigma_{12}}{\partial\omega}\left(A_{12}({\bf
k}, 0)+A_{21}({\bf k},0)\right)\right],
\end{eqnarray}
where $A_{11}$ is given in Eq.~\ref{A2A} and $A_{22}=A_{11}({\bf
k}\rightarrow{\bf k}+{\bf Q})$, i.e., $A_{22}$ is similar to
$A_{11}$, only the weights ($\alpha_{{\bf k}}^2$, $\beta_{{\bf
k}}^2$) are interchanged. We have used Eq.~\ref{BcVA} in the
calculation of Fig.~5. The constraint $\alpha_{{\bf
k}}^2+\beta_{{\bf k}}^2=1$ removes these SDW coherence factors
from the equation.  Neglecting the off-diagonal term proportional
to the small quantity $\partial\Sigma_{12}/\partial\omega$, the
specific heat expression becomes
\begin{eqnarray}\label{BcVN}
c_V&\approx &\frac{2}{3}\pi^2k_B^2T\sum_{{\bf k}}\left[
\delta(-Z_{\omega}^0E_{{\bf k}}^+)+\delta(-Z_{\omega}^0E_{{\bf
k}}^-)\right]
+ ...\nonumber\\
&\approx&\frac{2}{3}\pi^2k_B^2T\left[N^+(0)+N^-(0)\right]/Z_{\omega}^0.
\end{eqnarray}
Here, $N^{\pm}(0)=\sum_{{\bf k}} \delta(-E_{{\bf k}}^{\pm})$ is the
total density of states for the (U/L)MBs at the Fermi level in the
SDW state, but without renormalization (mean-field theory). For
both NCCO and LSCO, Eq.~\ref{BcVN} is an excellent approximation
to the exact expression of Eq.~\ref{BcVA}.
It is interesting to observe that in the final expression for $c_V$
(Eq.~\ref{BcVN}),
the self energy correction enters only through the renormalized DOS. Thus,
in the QP-GW model, $c_V$ has the same form as in conventional
Fermi-liquid theory.


\begin{thebibliography}{99}
\bibitem{foot3}$n(k)$ here is strictly the occupation density in the
reduced zone and not
the electronic momentum density, which can be measured via Compton
scattering (see, e.g Ref.~\onlinecite{Tanaka}) or
positron-annihilation (see, e.g. Ref.~\onlinecite{ft_posi1})
experiments.
\bibitem{shadow} Shadow bands have recently been observed by ARPES in
several cuprates.  However,
there are currently debates about whether they are consistent with $(\pi
,\pi )$\cite{chang,he}
or incommensurate order\cite{yang1,meng}, and whether they are related to
pseudogap
physics or parasitical structural superlattices.
\bibitem{foot2}The TB parmeters are taken to be doping independent in the
spirit of a
rigid band picture. A more realistic treatment of the doping
dependence of the electronic states (see,
e.g., Refs.~\onlinecite{ft_sophi1,ft_sophi3,ft_sophi4}) was not
undertaken. However, we expect the
rigid band model to be a good approximation for doping away from the Cu-O
planes.
\end{thebibliography}
\end{document}